\documentclass[showpacs,aps,prl,twocolumn,nofootinbib]{revtex4}
\usepackage{amsmath}
\usepackage{epsfig}
\usepackage{subfigure}
\usepackage{float}
\usepackage{verbatim} 
\usepackage{axodraw4j} 
\usepackage{pstricks}
\usepackage{color}
\usepackage{slashed}
\usepackage{multirow}
\usepackage{pstricks}
\usepackage{color}
\usepackage{booktabs}
\usepackage{subfigure}
\usepackage{epstopdf}

\newcommand{\ie}{{\it i.e.}}

\newcommand{\be}{\begin{equation}}
\newcommand{\ee}{\end{equation}}

\def\bsp#1\esp{\begin{split}#1\end{split}}

\newcommand{\tbn}[1]{Table~\ref{#1}}
\newcommand{\eqn}[1]{Eq.~(\ref{#1})}
\newcommand{\fig}[1]{Figure~\ref{#1}}
\newcommand{\figsc}[2]{Figures~\ref{#1},~\ref{#2}}
\newcommand{\ra}{\rightarrow}

\newcommand{\TeV}{\mbox{ ${\mathrm TeV}$}}

\begin{document}

\leftline{}
\rightline{CP3-12-XX}

\title{Effective field theory approach to the Higgs lineshape}

\author{Diogo Buarque Franzosi, Fabio Maltoni, Cen Zhang}
\affiliation{
Centre for Cosmology, Particle Physics and Phenomenology, 
Universit\'e Catholique de Louvain, B-1348 Louvain-la-Neuve, Belgium 
}

\begin{abstract}

The phenomenology of unstable particles, including searches and
exclusion limits at the LHC, depends significantly on its lineshape.
When the width of the resonance is large with respect to its mass, off-shell effects become relevant and the very same definition of width becomes non trivial.
Taking a heavy Higgs boson as an example, we propose a new
formulation to describe the lineshape via an effective field theory approach. Our method leads to amplitudes that are gauge invariant, respect  unitarity and can appropriately describe the lineshape of  broad resonances. The application of the method to the following relevant processes for the LHC phenomenology have been considered:  gluon fusion, vector boson scattering and $t\bar{t}$ production via weak boson fusion.

\end{abstract} 


\maketitle

\section{Introduction}

The CMS and ATLAS collaborations have announced the discovery of
a resonance around  $126$ GeV compatible with the Higgs boson predicted by the Standard Model (SM)~\cite{:2012an,:2012gu}.  To eventually confirm the discovery of the SM Higgs boson, it will be
necessary not only to measure  the strength and the structure of its couplings to the fermions 
and vector bosons of the SM, but also to exclude the existence of other heavier scalar states with similar properties.

The scalar sector of the SM is particularly simple, yet it does not provide any clue 
either on its origin, i.e. on the possible underlying dynamics, or on its stability.
Alternative models  that address these and other open questions often predict a richer
structure for the scalar sector, such as in supersymmetry (SUSY), 
Technicolor theories and models with extra dimensions.
Two-Higgs doublet models provide the simplest and most often studied extensions of the scalar sector of the SM.
 In all these cases, heavier scalar and/or pseudo scalar partners of the ``SM Higgs'' can be expected below the TeV scale.

The width of such heavy states, turns out to be sizable. 
A large width induces both a smearing and deformation of the
signal line shape as well as a sizable signal/background interference. For  a heavy SM-like Higgs, 
(e.g, $\Gamma\simeq$100 GeV for $m_H \simeq 550$ GeV), the narrow width approximation (NWA) has been 
shown to be untenable, possibly invalidating the currently set exclusion limits on the heavy Higgs 
and beyond~\cite{Passarino:2012ri,Kauer:2012hd}.

Going beyond the NWA, however, turns out not to be such an easy theoretical
task. The challenges are manifold. First, the most accurate predictions for the
signal cross sections, typically at the next-to-next-to-leading order (NNLO) in
QCD and at NLO in EW ,  assume a complete factorization between production and
decay, \ie, they employ the NWA, and the {\it a posteriori} inclusion of width
effects is not uniquely defined \cite{Goria:2011wa}.  Second, the very
definition of a width, which amounts to a resummation of a specific subset of
terms appearing at all orders in perturbation theory becomes problematic,
leading to possible violations of gauge symmetry as well as of unitarity
\cite{Argyres:1995ym}.

Currently, the most pragmatic and employed solution is the so-called complex
mass scheme (CMS) \cite{Denner:1999gp}. In short, it amounts to
analytically continue  the parameters entering the SM Lagrangian that are
related to the masses to complex values. Such scheme maintains gauge invariance
at all orders in perturbation theory and can be consistently employed in (N)NLO in
EW computations~\cite{Denner:2005fg,Actis:2008uh}. 
However, since a fixed complex pole is assumed for any
virtuality of the unstable particle, the resulting scattering amplitudes
violate unitarity and do not correctly describe the lineshape of broad resonances. 

An alternative to the CMS is the
fermion-loop scheme~\cite{Argyres:1995ym,Beenakker:1996kn}, which offers a solution for
the restoration of unitarity, but yet does not give a complete description of unstable particles because the width of a heavy Higgs is dominated by its decay into gauge bosons.
A consistent treatment of the bosonic contributions is possible in the framework
of the background field gauge \cite{Abbott:1980hw}, but this requires a
calculation of the complete radiative corrections at a fixed loop order. These two approaches
require greatly increase the complexity of the calculation. Other
suggested schemes for the treatment of unstable particles include the pole
scheme \cite{Stuart:1991xk}, the Seymour scheme \cite{Seymour:1995qg}, the
use of an effective Lagrangian including non-local interactions
\cite{Beenakker:1999hi,Beenakker:2003va}, and an approach based on
collinear effective field theory~\cite{Beneke:2003xh}.  

Summarizing, in general and especially for a heavy Higgs, one would like to be able to include a running
width in the propagator directly connected with the Higgs self-energy 
and at the same time to respect gauge invariance and unitarity.

In this work we  tackle the problem from an effective field theory (EFT) point
of view. We propose to systematically include width effects via a set  of gauge
invariant higher dimensional terms to the SM lagrangian, along the lines of what
was first proposed in Ref.~\cite{Beenakker:1999hi,Beenakker:2003va}. Such new
operators  systematically encapsulate higher  order terms coming from the
self-energy and naturally allow a running and physical  width for the Higgs in a
gauge invariant way.  As we will show in the following, our scheme is consistent
at higher orders and it can be considered a generalization of the CMS as it
reduces to it in the limit where the dependence on the virtuality of the Higgs
self-energy is neglected. 
 
\section{Setting up the stage}

The two-point Green's function for the Higgs boson is
\begin{equation}
\Delta_{H}(s)=s-m_{H,0}^2+\Pi_{HH}(s)\ ,
\end{equation}
where $m_{H,0}$ is the bare mass, and $\Pi_{HH}(s)$ is the Higgs self-energy.
In the conventional on-shell definition, the mass and width are given by
\begin{flalign}
&m_{H,OS}^2=m_{H,0}^2-{\rm Re} \Pi(m_{H,OS}^2)\ ,\\
&m_{H,OS}\Gamma_{H,OS}=\frac{{\rm Im} \Pi(m_{H,OS}^2)}{1+{\rm Re} \Pi'(m_{H,OS}^2)}\ .
\end{flalign}
These definitions become gauge-dependent at order $\mathcal{O}(g^4)$.

In order to avoid the divergence of the tree-level propagator $D(s)=i/(s-m_{H,OS}^2)$,
one performs the Dyson resummation to obtain
\begin{equation}
D(s)=\frac{i}{s-m_{H,OS}^2+im_{H,OS}\Gamma_{H,OS}}\ .
\end{equation}
To include the running effects of the width, one can further approximate the
propagator by
\begin{equation}
D(s)=\frac{i}{s-m_{H,OS}^2+i{\rm Im}\Pi(s)}\ ,
\end{equation}
where the imaginary part of $\Pi(s)$ is related to the Higgs-boson width. The
consistency of the above treatments of the Higgs propagator with the equivalence
theorem and unitarity has been discussed by Valencia and Willenbrock
\cite{Valencia:1990jp}.

Alternatively, as shown in a series of papers~\cite{Stuart:1991xk,Sirlin:1991rt}
a consistent, convenient and resilient definition of mass $\mu$ and width
$\gamma$ up to two loops, is obtained by setting $s_H \equiv  \mu^2 - i \mu
\gamma$ and then solving the implicit equation 
\begin{equation}
s_H - m_{H,0}^2 + \Pi_{HH}(s_H) =0
\end{equation}
in terms of $s_H$. This gives a gauge independent definition to all orders
\cite{Gambino:1999ai} (independent of the gauge choice present in the
computation of  $\Pi_{HH}(s_H)$) and in addition avoids unphysical threshold
singularities \cite{Kniehl:2002wn}.

The above definition is also consistent with the use of the CMS. In this scheme
the propagator is $\Delta_{H}^{-1}(s)=s- s_H$. By definition this approach can
give a good approximation of the full propagator 
\begin{equation}\label{eq:prop}
\Delta_{H}^{-1}(s)=\frac{1}{s-s_H+\Pi^R_{HH}(s)}
\end{equation}
only close to the pole or, equivalently, for a small width, $\gamma/ \mu\ll 1$.
Here $\Pi^R_{HH}(s)$ is the renormalized self energy, satisfying the following
renormalization conditions:
\begin{equation}
\Pi^R_{HH}(s_H)=0\ ,\quad \Pi'^R_{HH}(s_H)=0\ .
\end{equation}

A natural improvement would consist in including  the full resummed propagator
in explicit calculations. This, however, leads to gauge violation already at the
tree level.  The reason being that in perturbation theory gauge invariance is
guaranteed order by order while the presence of a  width implies the resummation
of  a specific subset of higher order contributions, the self-energy
corrections.  This results in a mixing of different orders of perturbation
theory. In particular, the following issues need to be addressed:\\
\begin{enumerate}
\item In general $\Pi_{HH}(s)$ explicitly depends on the gauge-fixing parameter
  (GFP).  To resum the self-energy correction to all orders, $\Pi_{HH}(s)$ must
  be extracted in a physically meaningful way.
\item The resummed propagator spoils the gauge cancellation among different
  diagrams, and eventually leads to the violation of Goldstone-boson equivalence
  theorem and unitarity bound.
\end{enumerate}
\begin{figure*}[t!]
  \includegraphics[width=1.5\columnwidth]{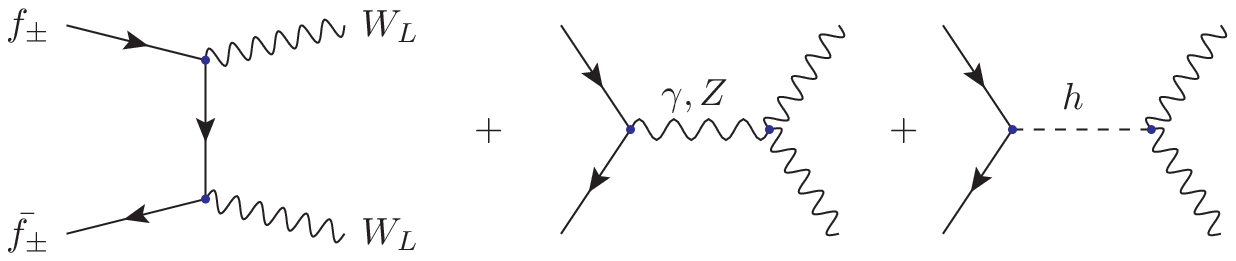}
  \caption{\label{fig:1} Diagrams contributing to fermion anti-fermion scattering into longitudinal $W$'s. 
}
\end{figure*}
Both issues can be tackled by the so-called Pinch Technique (PT)
\cite{Cornwall:1981zr,Cornwall:1989gv,Papavassiliou:1989zd,Degrassi:1992ue}. In
the PT framework, a modified one-loop self-energy for the Higgs boson can be
constructed by appending to the conventional self energy additional
propagator-like contributions concealed inside vertices and boxes.  For the
application of PT in resonant transition amplitude, and in particular, the
extraction of a physical self energy, we refer to the work of
Refs.~\cite{Papavassiliou:1995fq,Papavassiliou:1996zn,Papavassiliou:1995gs,Papavassiliou:1997fn,Papavassiliou:1998pb}.

The modified self-energy correction for the Higgs is GFP-independent, and
reflects properties generally associated with physical observables.  At the one loop
level, we have the following expressions
\cite{Papavassiliou:1997fn,Papavassiliou:1997pb}
\begin{widetext}
\begin{flalign}\label{eq:PiPT1}
\Pi^{(WW)}_{HH}(s)&=\frac{\alpha_W}{16\pi}\frac{m_H^4}{m_W^2}\left[1+4\frac{m_W^2}{m_H^2}-4\frac{m_W^2}{m_H^4}(2s-3m_W^2)\right]
B_0(s,m_W^2,m_W^2)\ ,
\\
\Pi^{(ZZ)}_{HH}(s)&=\frac{\alpha_W}{32\pi}\frac{m_H^4}{m_W^2}\left[1+4\frac{m_Z^2}{m_H^2}-4\frac{m_Z^2}{m_H^4}(2s-3m_Z^2)\right]
B_0(s,m_Z^2,m_Z^2)\ ,
\\
\Pi^{(ff)}_{HH}(s)&=\frac{3\alpha_W}{8\pi}\frac{m_f^2}{m_W^2}\left(s-4m_f^2\right)B_0(s,m_f^2,m_f^2)\ ,
\\
\Pi^{(HH)}_{HH}(s)&=\frac{9\alpha_W}{32\pi}\frac{m_H^4}{m_W^2}B_0(s,m_H^2,m_H^2)\ ,
\label{eq:PiPT4}
\end{flalign}
\end{widetext}
where the superscripts denote the contributions from the $W$, $Z$, fermions and
Higgs loops, and
\begin{flalign}
B_0(p^2,m_1^2,m_2^2)&\nonumber\\
\equiv
(2\pi\mu)^{4-d}&\int\frac{d^dk}{i\pi^2}\frac{1}{\left(k^2-m_1^2\right)\left[(k+p)^2-m_2^2\right]}
\end{flalign}
is the normal Passarino-Veltman function \cite{'tHooft:1978xw}. These results
are indepent of the GFP. Note that the expressions in
Eqs.~(\ref{eq:PiPT1}-\ref{eq:PiPT4}) coincide with the $\xi=1$ result obtained in
the background-field gauge
\cite{Denner:1994nn,Hashimoto:1994ct,Papavassiliou:1994yi}.

In addition, the gauge cancellation among different amplitudes can be restored,
by including certain vertex corrections obtained via the PT~
\cite{Papavassiliou:1996fn,Papavassiliou:1997pb}.  This is because in this
framework the Green's functions  satisfy the tree-level-like Ward identities
(WI), which are crucial for ensuring the gauge invariance of the resummed
amplitude.

As an example, let us consider the Higgs-mediated part of same helicity fermion
scattering into longitudinal $W$'s, $f_\pm\bar{f}_\pm \to W^+_LW^-_L$. There are
contributions from $s$-channel and $t$-channel diagrams, as is shown in
\fig{fig:1}.  The contributions from $t$-channel and Higgs diagram to the
amplitudes coming from longitudinal components of the $W$'s and same helicity
fermions (in the high energy limit) read \footnote{Here we assume
  $\epsilon^\mu_{1,2}\approx k^\mu_{1,2}/m_W$ at high energy region.}
\begin{flalign}\label{eq:ffww}
{\cal M}^L_{h} &\equiv {\cal M}_s^{\mu\nu}\frac{k_{1\mu} k_{2\nu}}{m_W^2}\nonumber\\
&=\frac{-igm_f}{2m_W}\bar{v}(p_2)u(p_1)\frac{i}{\Delta_{H}(s)}\Gamma^{HWW,\mu\nu}(q,k_1,k_2)\frac{k_{1\mu} k_{2\nu}}{m_W^2}
\nonumber\\
{\cal M}^L_{t}&\equiv {\cal M}_{t,Z}^{\mu\nu}\frac{k_{1\mu} k_{2\nu}}{m_W^2}\nonumber\\
&=-\frac{ig^2 m_f }{4m_W^2}\bar{v}(p_2)u(p_1)+\cdots
\end{flalign}
where $\Gamma_{\mu\nu}^{HWW}(q,k_1,k_2)$ is the $HW^+W^-$ vertex.  The ellipsis
in ${\cal M}^L_{t}$ denotes terms that are not related to the Higgs exchange
diagram.  These terms come from the contribution of opposite helicity fermions,
and are supposed to cancel the bad high-energy behavior of the $\gamma/Z$
mediated diagrams.
 
Without the Higgs contribution  ${\cal M}^L$ grows with energy and eventually
violates unitarity.  The cancellation of the bad high-energy behavior of each
amplitude, and the equivalence theorem, are guaranteed by the following WI: 
\begin{flalign}
\label{eq:wi1}
&k_+^\mu k_-^\nu\Gamma_{\mu\nu}^{HWW}(q,k_+,k_-)=\nonumber\\
&-m_W^2\Gamma^{H\phi^+\phi^-}(q,k_+,k_-)+\frac{igm_W}{2}\Delta_{H}(q^2)\,,
\end{flalign}
where $\phi^{\pm}$ are Nambu-Goldstone bosons. Only the leading terms at high energy
are included. The relation above explicitly shows that the inclusion of higher
order terms in the imaginary part of $\Delta_{H}(q^2)$ has to be related to the
EW corrections of $\Gamma^{HWW}_{\mu\nu}$ and three scalar vertex. Only if both
$\Delta_{H}(s)$ and $\Gamma^{HWW}_{\mu\nu}$ are computed in one-loop via the PT,
then the WI remains valid, and the gauge-cancellation, as well as the
equivalence theorem, are not spoiled. Besides, ${\cal M}^L_{t,Z}$ is not
affected by the Higgs width and therefore the tree-level relations can be used.
Thus the resummed propagator can be consistently included with the one-loop
correction to $\Gamma^{HWW}_{\mu\nu}$ via the PT.

Even though correct, the solution  outlined above for $f\bar{f}\to W^+_LW^-_L$
is not  a general one.  In $W^+_LW^-_L\rightarrow Z_LZ_L$, for example, it is
not sufficient to include only the $HWW$ and $HZZ$ corrections.  The
triple and quartic vector-boson vertices at one-loop are also required to cancel the bad
high-energy behavior of the Higgs-mediated amplitude and the overall procedure
of analyzing the full set of WI's becomes more and more involved. The goal of
this work is to present a simple method to generate the needed corrections to
the vertices and propagators so that the WI's are automatically satisfied and
unitarity automatically ensured.

\section{The EFT approach}

As explained above, we aim at finding a systematic approach to improve the Higgs propagator
without breaking either gauge invariance or unitarity. In other words we are
looking for a mechanism that guarantees the constraints imposed by the WI 
to be satisfied at any order in perturbation theory.

At one loop, the full calculation via the PT  certainly provides an exact
solution valid at NLO.  The challenge is to achieve the same keeping the
calculation at leading order, including only the necessary ingredients coming
from NLO and resumming them into the propagator via a Dyson-Schwinger approach.
The idea is to associate the corrections to an ad hoc constructed
gauge-invariant operator and match the operator to the one-loop two-point
function $\Delta_{H}(s)$ calculated via the PT.  In so doing one aims at
obtaining  the exact resummed propagator already at the leading order and, {\it
at the same time},  the  interactions modified to automatically satisfy the
WI's.  The latter desired result ensures the gauge-invariance of the amplitudes,
and it can be considered as an approximation to a full one-loop calculation in
PT.

To this aim, we consider the Taylor expansion of the function $\Pi(s)=\Pi^R_{HH}(s)$
\begin{equation}
\Pi(s)=\sum_{i=0}^\infty c_i s^i \,,
\end{equation}
where $c_i$ are dimensionful constants and,  as first attempt, we add the
following infinite set of operators to the Lagrangian:
\begin{flalign}\label{eq:op1}
  {\cal O}_\Pi=&\sum_{i=0}^\infty c_i \phi^\dagger(-D^2)^i\phi
\nonumber\\
\equiv&\phi^\dagger \Pi(-D^2)\phi
\end{flalign}
where $\phi$ is the Higgs doublet, and $D^\mu$ is the covariant derivative.  It
is straightforward to check that  ${\cal O}_\Pi$ modifies the Higgs propagator as
desired: the two $\phi$'s contribute two Higgs fields, and each $-D^2$
contributes an $s$ leading to 
\begin{equation}\label{eq:PiHH}
\Pi(s)=\Pi_{HH}^R(s)\,,
\end{equation}
as desired. Note that in principle, ${\cal O}_\Pi$ is a non-local operator, yet by
expanding it, we re-express it in terms of an infinite series of local
operators.\footnote{In general,  inclusion of higher-order derivatives in the Lagrangian leads
to very peculiar quantum field theories, aka Lee-Wick theories, see \cite{Grinstein:2007mp} for a recent analysis and references. 
As we are going to see later, in our approach, we only use the imaginary part of $\Pi(s)$ and therefore the real part of the propagator is not affected.
}

We remark that while very similar in spirit , our approach differs
from that of Ref.~\cite{Beenakker:1999hi}:
the operator chosen there does not contain gauge fields, and it is therefore not
sufficient to restore the gauge cancellation and fix the bad high-energy
behavior in vector-vector scattering.
\begin{figure*}[t]
\begin{center}
  \includegraphics[width=1.9\columnwidth]{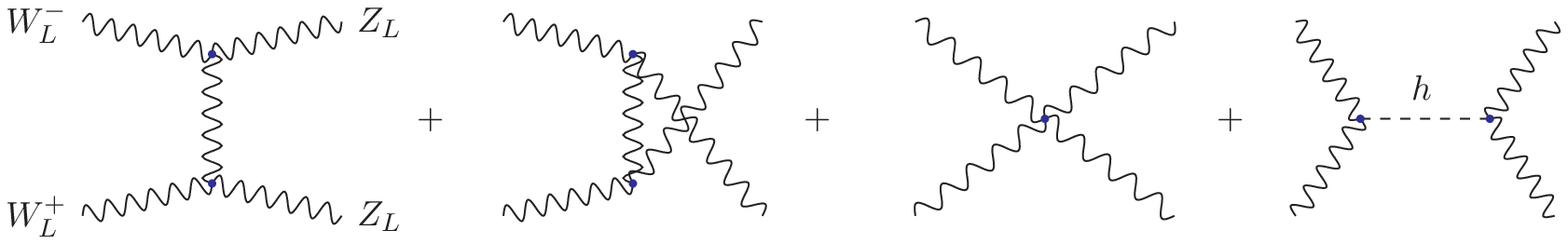}
\end{center}
\caption{Diagrams contributing to $W^+_LW^-_L\rightarrow Z_LZ_L$
\label{fig:2} }
\end{figure*}

Eq.~(\ref{eq:op1}) leads to the correct expression for the propagator. However,
the first term $\Pi(0)\phi^\dagger\phi$ in the expansion corresponds to a
tadpole contribution. This can be avoided if this term  is replaced by 
\[
\Pi(0)\phi^\dagger\phi
\to
\frac{\Pi(0)}{2v^2}\left[(\phi^\dagger\phi)-\frac{v^2}{2}\right]^2\,,
\]
\ie, the Higgs-self interaction is suitably modified. As one can easily  check,
such a modification leaves  the relation of Eq.~(\ref{eq:PiHH})  unchanged.  The
final form of the operator, which we dub ${\cal \tilde{O}}_\Pi$, is   
\begin{equation}\label{eq:O}
  {\cal \tilde{O}}_\Pi=\phi^\dagger[\Pi(-D^2)-\Pi(0)]\phi+\frac{\Pi(0)}{2v^2}\left[(\phi^\dagger\phi)-\frac{v^2}{2}\right]^2\,.
\end{equation}
The addition  of this operator to the SM leads to several changes,  which we now
consider in detail.  First of all, by construction, it gives rise to the
propagator in Eq.~(\ref{eq:prop}), and a resummed propagator with the full
one-loop self energy via the PT at tree level is obtained.  Second it leads to
modifications of the other interactions, in such a way that gauge invariance is
maintained.  For example, the $W$ and $Z$ two-point functions are modified by
the addition of 
\begin{flalign}
i\Delta\Pi^{\mu\nu}_{WW}(q^2)=&i\left(\frac{gv}{2}\right)^2\left[\Pi'(0)g^{\mu\nu}+\Pi''(q^2)q^\mu q^\nu\right]\nonumber
\\
i\Delta\Pi^{\mu\nu}_{ZZ}(q^2)=&i\left(\frac{gv}{2c_W}\right)^2\left[\Pi'(0)g^{\mu\nu}+\Pi''(q^2)q^\mu q^\nu\right]\nonumber
\end{flalign}
where $v$ is the Higgs vev, and
\begin{equation}
\Pi'(x)\equiv \frac{\Pi(x)-\Pi(0)}{x},\qquad  \Pi''(x)\equiv \frac{\Pi'(x)-\Pi'(0)}{x} \,.
\end{equation}
The values for the $W$ and $Z$ masses are  shifted
\begin{flalign}
&m_W^2=\left(\frac{gv}{2}\right)^2(1+\Pi'(0))\nonumber \\
&m_Z^2=\left(\frac{gv}{2c_W}\right)^2(1+\Pi'(0))\,,
\label{eq:EFTmass}
\end{flalign}
as well as the propagators 
\begin{equation}
\frac{i}{q^2-m_{W,Z}^2}\left[
- g^{\mu\nu}+\frac{\left(1+\frac{m_{W,Z}^2\Pi''(q^2)}{1+\Pi'(0)}\right)q^\mu q^\nu}{m_{W,Z}^2+q^2\frac{m_{W,Z}^2\Pi''(q^2)}{1+\Pi'(0)}}
\right]\,.
\label{eq:EFTprop}
\end{equation}

Let us first consider $f\bar{f}\to W_L^+W_L^-$ in the EFT approach. The operator
modifies the $HW^+W^-$ and the $Hf\bar{f}$ interactions.  The combined effect is
a factor of $1+\Pi'(s)$. Therefore in this process the EFT approach is
equivalent to the following substitution of the Higgs propagator:
\begin{equation}
\Delta_{H}^{-1}(s)=\frac{1+\Pi'(s)}{s-s_H+\Pi(s)}\ ,
\end{equation}
which behaves like $1/s$ at large energy, and therefore exactly cancels the
high-energy behavior from ${\cal M}^L_{t}$.  It is also interesting to note
that, if $\Pi(s)$ has a linear dependence on $s$, \ie,
\begin{equation}
\Pi(s)=i(s-\mu^2)\frac{\gamma}{\mu}\,,
\end{equation}
the above equation becomes
\begin{equation}
\Delta_{H}^{-1}(s)=\frac{1+i\frac{\gamma}{\mu}}{s-\mu^2+is\frac{\gamma}{\mu}}\ ,
\end{equation}
and the EFT approach coincides with the scheme proposed by Seymour
\cite{Seymour:1995qg}.  This makes sense because in the Seymour scheme the
vector boson pair self energy also has a linear dependence on $s$. In our scheme
we see that the numerator of Seymour's propagator comes from the modified
$HW^+W^-$ vertex, as required by the WI.

We now turn to  vector-vector scattering and in particular to
$W^+_LW^-_L\rightarrow Z_LZ_L$.  This process features a pure gauge and a
Higgs-mediated $s$-channel contribution,  Fig.~\ref{fig:2}.  Both contributions
do contain terms that grow as $s$ at high energy, whose cancellation is
guaranteed by gauge-invariance. To calculate $W^+_LW^-_L\rightarrow Z_LZ_L$
amplitude in the EFT we need to extract the Feynman rules from ${\cal \tilde{O}}_\Pi$,
\ie, the contributions that need to be added to the usual SM rules.  This is
straightforward and gives (all momenta incoming):
\begin{widetext}
\begin{equation}
\begin{array}{ll}
 H(q)W^{+\mu}(k_1)W^{-\nu}(k_2)   
& ig\frac{m_W}{\sqrt{1+\Pi'(0)}}\Pi'(q^2)g^{\mu\nu}+\cdots  \\
 Z^\mu(k_1)W^{+\nu}(k_2)W^{-\rho}(k_3) 
& i\frac{g}{c_W}\frac{m_W^2}{1+\Pi'(0)}s_W^2\left[
\Pi''(k_3^2)g^{\mu\nu}k_3^\rho-\Pi''(k_2^2)g^{\mu\rho}k_2^\nu\right]+\cdots \\
Z^\mu(k_1)Z^\nu(k_2)W^{+\rho}(k_3)W^{-\sigma}(k_4)\hspace*{1cm}
&  ig^2\frac{m_Z^2}{1+\Pi'(0)}\left[\Pi''(s)g^{\mu\nu}g^{\rho\sigma} +s_W^4(\Pi''(t)g^{\mu\rho}g^{\nu\sigma}+\Pi''(u)g^{\mu\sigma}g^{\nu\rho})\right]+\cdots \\
H\phi^+\phi^-,\ H\phi^0\phi^0 
& -i\frac{g[m_H^2-\Pi(0)]}{2m_W}\sqrt{1+\Pi'(0)}\\
\phi^+\phi^-\phi^0\phi^0
& -i\frac{g^2[m_H^2-\Pi(0)]}{4m_W^2}[1+\Pi'(0)]
\end{array}
\label{eq:fr}
\end{equation}
\end{widetext}
where ellipsis denotes terms vanishing on shell 
and $s=(k_1+k_2)^2$, $t=(k_1+k_3)^2$, and $u=(k_1+k_4)^2$.  These Feynman rules are sufficient to calculate
both $W^+_LW^-_L\to Z_LZ_L$ and $\phi^+\phi^-\to \phi^0\phi^0$. At the leading order in
$\frac{m_W^2}{s}$ and $\frac{m_W^2}{m_H^2}$, we find for $W^+_LW^-_L\to Z_LZ_L$,
\begin{flalign}
  &{\cal M}_H^{LLLL}=-\frac{ig^2}{4m_W^2}\frac{s^2[1+\Pi'(s)]^2}{[s-m_H^2+\Pi(s)][1+\Pi'(0)]}
\\
&{\cal M}_{\rm gauge}^{LLLL}=\frac{ig^2}{4m_W^2}s\frac{1+\Pi'(s)}{1+\Pi'(0)}
\end{flalign}
and for $\phi^+\phi^-\to \phi^0\phi^0$,
\begin{flalign}
&{\cal M}_G=-\frac{ig^2}{4m_W^2}\frac{s+\Pi(s)-\Pi(0)}{s-m_H^2+\Pi(s)}\left[m_H^2-\Pi(0)\right]\left[1+\Pi'(0)\right]
\end{flalign}
so that
\begin{flalign}
  {\cal M}_H^{LLLL}+{\cal M}_{\rm gauge}^{LLLL}&=-\frac{ig^2}{4m_W^2}\frac{s+\Pi(s)-\Pi(0)}{s-m_H^2+\Pi(s)}\frac{m_H^2-\Pi(0)}{1+\Pi'(0)}\nonumber \\
&= \frac{{\cal M}_G}{[1+\Pi'(0)]^2}\,.
\end{flalign}
As expected, ${\cal M}_H^{LLLL}+{\cal M}_{\rm gauge}^{LLLL}$ does not grow with
$s$ and the equivalence theorem is recovered, up to a factor $[1+\Pi'(0)]^2$,
which exactly amounts to the wave function renormalization of the Goldstone
fields.

An interesting feature of our approach is that in the limit where the dependence
of $\Pi(s)$ on $s$ is neglected,  $\Pi(s)\equiv\Pi$ is a constant, then
$\Pi'(s)=\Pi''(s)=0$.   The only effect of the operator is a shift in $\lambda$,
the coupling of the Higgs-boson self interaction.  If $m_H$ is the on-shell
mass, this amounts to the replacement
\begin{equation}
m_H^2\to m_H^2-\Pi \,,
\end{equation}
\ie, given that $\Pi$ can be a complex number, it is equivalent to the CMS. 

The advantage of the EFT approach is the possibility of using ``arbitrary'' functional
form of the self energy.  We have shown that with special choices of $\Pi(s)$,
the EFT approach can reduce to the Seymour scheme and the CMS scheme in certain
cases. For example, there is no need for spurious non-zero width for
$t$-channel propagators as this can be easily imposed by always maintaining  gauge invariance.
Finally, we note that despite the restoration of gauge-invariance and
equivalence theorem is a general feature of our approach, one has to be careful in 
choosing the appropriate operator. For example, 
the following operator
\begin{equation}\label{eq:op2}
  {\cal O}'_\Pi=\frac{1}{2v^2}\left(\phi^\dagger\phi-v^2\right)\Pi(-\partial^2)\left(\phi^\dagger\phi-v^2\right)
\end{equation}
introduced in Ref.~\cite{Beenakker:1999hi},
 gives rise to the correct self energy and the resummed propagator, but it does not
modify the gauge contribution, so in $W^+_LW^-_L\to Z_LZ_L$ the gauge
cancellation between the $s$-channel Higgs-mediated amplitude and the gauge
amplitude is not restored. On the other hand, it modifies the Goldstone
amplitude in a way so that the equivalence theorem is satisfied. As a result,
both $W^+_LW^-_L\to Z_LZ_L$ and $\phi^+\phi^-\to \phi^0\phi^0$ have bad high-energy
behavior, and eventually break unitarity bounds.  In general, adding higher
dimensional operators to the Lagrangian leads to unitarity violation at some
scale.  We are going to show in the next sections that the operators we use do not have this problem. 

Though the above operator $\mathcal{O}'_\Pi$ solely does not treat the $HZZ$ and
$HW^+W^-$ correctly at high energy, when combined with $\mathcal{O}_\Pi$, we can
adjust them in a certain way to improve this method. We will discuss this in
the following sections.

\section{Unitarity}
\label{sec:unitarity}

Adding operators of dimension $n>4$ to the SM Lagrangian
\begin{equation}
\mathcal{L}_{EFT} = {\cal L}_{SM} + \sum_i c_i \frac{{\cal O}_i[n]}{\Lambda^{n-4}}\,,
\end{equation}
 is equivalent to recast the SM in terms of an effective field theory valid up
 to scales of order $\Lambda$ \cite{Weinberg:1978kz}, beyond which the theory is
 not unitary. It is therefore mandatory to check whether this is the case for
 the operator ${\cal \tilde{O}}_{\Pi}$.  In fact,  as we will see in the following
 section, a consistent perturbation theory implies that the same operator needs
 to also appear as a counterterm  at higher orders. Overall  we do not modify
 the theory and our procedure amounts to a reorganization of the perturbative
 expansion. However, we still need to make sure that neither unitarity is  violated
 nor double counting happens  at any given order in the perturbation theory. In
 this section we consider the first of these issues by showing that in sample
 calculations, $f\bar{f}\to VV$ and $VV\to VV$, at tree-level the operator in
 Eq.~(\ref{eq:O}) does not break unitarity at large energy. 

In $f_\pm\bar{f}_\pm\to VV$ the change in $HVV$ vertex cancels
the change in $H$ propagator at high $s$, independently of the helicities of
$VV$, so the $s$-channel Higgs diagram does
not lead to any bad high-energy behavior. The scattering of opposite helicity
fermions does not entail the $s$-channel Higgs diagram and is the same as in the
SM.

As we have already verified, in $W^+W^-\to ZZ$  the longitudinal amplitude does
not break unitarity, because the modification to the corresponding Goldstone
interaction is finite ($m_H^2-\Pi(0)$ and $1+\Pi'(0)$).  We now check the
transverse amplitude $++\to --$, $00\to ++$, $++\to 00$, $++\to ++$, in the limit
\begin{flalign}
  s\sim |t|\sim |u|\gg m_W^2,\quad m_H^2\gg m_W^2\,.
\end{flalign}
(Note that $+-$, $+0$,
$-0$ configurations do not feature a Higgs in the $s$-channel and therefore are
left unchanged.) An explicit calculation for $W^+_+W^-_+\to Z_-Z_-$ gives
\begin{flalign}
&{\cal M}_H^{++--}={\cal M}_H^{LLLL}\frac{4m_W^4}{s^2c_W^2}+\mathcal{O}(m_W^4)\\
&{\cal M}_{\rm gauge}^{++--}={\cal M}_{\rm gauge}^{LLLL}\frac{4m_W^4}{s^2c_W^2}+\mathcal{O}(m_W^4)\,,
\end{flalign}
where $M^{LLLL}$ indicates the amplitude with four longitudinal vectors.  For
$W^+_LW^-_L\to Z_+Z_+$ we obtain
\begin{flalign}
&{\cal M}_H^{LL++}={\cal M}_H^{LLLL}\frac{-2m_W^2}{sc_W^2}+\mathcal{O}(m_W^2)\\
&{\cal M}_{\rm gauge}^{LL++}={\cal M}_{\rm gauge}^{LLLL}\frac{-2m_W^2}{sc_W^2}+\mathcal{O}(m_W^2)\,,
\end{flalign}
and for
$W^+_+W^-_+\to Z_LZ_L$ we obtain
\begin{flalign}
  &{\cal M}_H^{++LL}={\cal M}_H^{LLLL}\frac{-2m_W^2}{s}+\mathcal{O}(m_W^2)\\
  &{\cal M}_{\rm gauge}^{++LL}={\cal M}_{\rm gauge}^{LLLL}\frac{-2m_W^2}{s}+\mathcal{O}(m_W^2)\,,
\end{flalign}
These results vanish faster than the longitudinal amplitude at large $s$.
Finally for $W^+_+W^-_+\to Z_+Z_+$, we obtain
\begin{flalign}
{\cal M}_H^{++++}&= -{ig^2m_Z^2}\frac{[1+\Pi'(s)]^2}{[s-m_H^2+\Pi(s)][1+\Pi'(0)]}
+\mathcal{O}(m_W^4) \nonumber\\
&\approx\frac{-{ig^2m_Z^2}}{1+\Pi'(0)}\Pi''(s)\sim s^{-1}\quad\mbox{at large $s$,}
\\
{\cal M}_{\rm gauge}^{++++}&=i8g^2c_W^2\frac{s^2}{4tu}+\mathcal{O}(m_W^2)\,.
\end{flalign}
so at large energy the inclusion of $\mathcal{\tilde{O}}_\Pi$ does not lead to
any bad high energy behavior.

\section{The EFT approach at higher orders}

Starting at order $\alpha_W$, the operator ${\cal\tilde{O}}_{\Pi}$ is allowed in any
leading order computation. At next-to-leading order in EW interactions, however,
this is not necessarily consistent and possibly leads to double counting. In
this section we argue that this is not  a fundamental problem and can be dealt
with by simply subtracting the same operator in a NLO as a counterterm, in full
analogy to the procedure used in the CMS \cite{Denner:2005fg}.

In the CMS, an imaginary part  is added to the real mass, and then subtracted as
counterterm at NLO.  One can prove that this procedure does not spoil the WI's,
despite the fact that only a special class of higher order terms is resummed. As
the EFT approach can be viewed as a generalization to the CMS, the same approach
can be followed.  The operator ${\cal\tilde{O}}_{\Pi}$ corresponds to the imaginary part
of the mass. It includes some of the higher-order contribution, and provides an
improved solution to the WI's. It enters the resummed propagator and other
Feynman rules, and  needs to be subtracted at higher orders. The main difference
is that in the CMS the propagator describes an unstable particle with a fixed
width, while in the EFT approach one can resum an arbitrary part of the
self-energy correction.  This difference may be important when the width of the
unstable particle is large, as in the case of a heavy Higgs, and the actual
functional form of $\Pi(s)$ becomes important.

In the pole-mass renormalization scheme, the two-point function of the Higgs can
be written as
\begin{equation}\label{eq:2point}
\Delta_{H}(s)=s-s_H+\Pi^R_{HH}(s)
\end{equation}
where $s_H$ is the pole and $\Pi^R_{HH}(s)$ is the one-loop PT self-energy
correction renormalized in the pole-mass scheme. We can now define the EFT
approach by adding the operator in Eq.~(\ref{eq:O}) and subtracting it as a
counterterm:
\begin{equation}
\mathcal{L}_{SM} \to\mathcal{L}_{SM} +{\cal\tilde{O}}_{\Pi}-{\cal\tilde{O}}_{\Pi}\,.
\end{equation}
In so doing the theory is exactly the same as before. Now Eq.~(\ref{eq:2point})
can be rewritten as
\begin{flalign}
\Delta_{H}(s)=&s-s_H+\Pi(s)\nonumber\\
&+[\Pi^R_{HH}(s)-\Pi(s)]
\end{flalign}
where the first line starts at leading order, while the second line starts at
order $\alpha_W$.  The EFT approach then amounts to choose $\Pi(s)$ in a way to
capture the important part of (if not all of) $\Pi^R_{HH}(s)$, so that this part
of the self-energy correction is included at the leading order, and will be
resummed.  In practice, one does not have to choose the exact PT self energy, and
gauge invariance is always guaranteed.  In particular, choosing $\Pi(s)=0$
corresponds to the CMS scheme.

In our scheme, EW NLO calculations are obviously more involved. The resummed
propagator (\ref{eq:prop}) and the modified Feynman rules do require extra work.
However, one can also always  employ a standard CMS at NLO and only include the
full propagators and vertices in the LO result. In this way we can consistently have
leading order calculated in the EFT approach, and NLO in CMS but with
counterterms from ${\cal\tilde{O}}_{\Pi}$.

\section{Improved operator}

\label{sec:improve} In the EFT approach, the exact functional
form of the $\Pi(s)$ is somewhat arbitrary. Ideally, one would like to use the imaginary part of the PT
one-loop self energy, as given in Eq.~(\ref{eq:PiPT1})-(\ref{eq:PiPT4}).
Unfortunately, in $VV\to VV$  this choice produces an unphysical excess at low energy. 
As mentioned above, such an excess can be avoided by incorporating the operator
$\mathcal{O}'_\Pi$ in \eqn{eq:op2}.

The reason for the excess is that the operator does not correctly describe the
1-loop $HVV$ vertices at low energy. This can be traced back to the WI in
Eq.~(\ref{eq:wi1})\footnote{In fact the r.h.s has
  additional terms
  $igm_Z\left[\Pi_{\phi^+\phi^-}(k_1^2)+\Pi_{\phi^+\phi^-}(k_2^2)\right]/2c_W$.
They are zero for this process because we consider only the imaginary part.}.  The operator
${\cal \tilde{O}}_\Pi$ in \eqn{eq:O} does not modify $\Gamma^{H\phi^+\phi^-}$,
hence, the equality relation is completely satisfied by the $HWW$ vertex. At
high energy this arrangement is satisfactory because we do not expect
contributions from the Goldstone vertex. However, at low energy, this vertex
turns out to be relevant and thus, modifying only the $HWW$ vertex gives rise to
a bad behavior at low energy. On the other
hand, the operator ${\cal O'}_\Pi$ (\eqn{eq:op2}) satisfies the equation with
the saturation of the Goldstone-Higgs vertex, yet fails at describing the high
energy behavior of weak boson scattering for example. The solution for this
problem relies on the construction of a suitable  combination of these two operators. 

Let us rewrite them in the following form:
\begin{flalign}
  {\cal O}_{\Pi_1}=\phi^\dagger\Pi_1(-D^2)\phi \,.
  \label{eq:opi1}
\end{flalign}
Here we consider the operator of \eqn{eq:O} in the simpler form, \eqn{eq:op1}.
This is justified by the fact that in practice we have to work only with the
imaginary part of the self energy, then $\Pi(0)=0$ \footnote{
  Also because in the end we can always shift a constant part from
  $\mathcal{O}_{\Pi_1}$ to ${\cal O'}_{\Pi_2}$.
}. On the other hand, the operator proposed by Beenakker \textit{et al.} \cite{Beenakker:1999hi} (\eqn{eq:op2}),
can be written as:
\begin{flalign}
  {\cal O}'_{\Pi_2}=\frac{1}{2v^2}\left(\phi^\dagger\phi-v^2\right)\Pi_2(-\partial^2)\left(\phi^\dagger\phi-v^2\right)\,.
  \label{eq:opi2}
\end{flalign}

To determine $\Pi_1$ and $\Pi_2$, we focus on the $HZZ$ vertex and the
$H\phi^0\phi^0$ vertex.  The inclusion of ${\cal O}_{\Pi_1}$ and ${\cal
O}'_{\Pi_2}$ modifies these vertices, respectively:
\begin{flalign}
  k_1^\mu k_2^\nu\Gamma_{HZZ,\mu\nu}(q,k_1,k_2)
  &=i\frac{gm_Z}{2c_W}\Pi_1(q^2)\\
  \Gamma_{H\phi^0\phi^0}(q,k_1,k_2)
  &=i\frac{g}{2m_W}\Pi_2(q^2)\,,
\end{flalign}
where $k_1$ and $k_2$ are the momenta of $Z^\mu Z^\nu$ and $\phi^0\phi^0$.
Now consider the following Ward Identity:
\begin{flalign}
\label{eq:wi2}
&k_1^\mu k_2^\nu\Gamma_{\mu\nu}^{HZZ}(q,k_1,k_2)=\nonumber\\
&-m_Z^2\Gamma^{H\phi^0\phi^0}(q,k_1,k_2)+\frac{igm_Z}{2c_W}\Delta_{H}(q^2)\,.
\end{flalign}
This equation implies
\begin{equation}
\Pi(q^2) = \Pi_1(q^2) + \Pi_2(q^2)\,.
\end{equation}
We now need to find out the expressions for $k_1^\mu
k_2^\nu\Gamma_{HZZ,\mu\nu}(q,k_1,k_2)$ and $\Gamma_{H\phi^0\phi^0}(q,k_1,k_2)$,
or equivalently, $\Pi_1(q^2)$ and $\Pi_2(q^2)$. Then the combined operator ${\cal
O}_{\Pi_1}+{\cal O}'_{\Pi_2}$ should reproduce all three terms in \eqn{eq:wi2}
correctly. To this end, we calculate the absorptive part of both $k_1^\mu
k_2^\nu\Gamma_{HZZ,\mu\nu}(q,k_1,k_2)$ and $\Gamma_{H\phi^0\phi^0}(q,k_1,k_2)$
at one loop with PT. The results obtained are:
\begin{flalign}
  \Pi_{1,2}(s)=\sum_{XX}\Pi_{1,2}^{XX}(s)
\label{eq:pi12}
\end{flalign}
where $XX$ are summed over $WW$, $ZZ$, $tt$ and $HH$, and

\begin{widetext}
  \begin{flalign}
  \Pi_1^{WW}(s)=&
  -\frac{\alpha_W}{4\pi}\left[2sB_0(s,m_W^2,m_W^2)+(s-2m_Z^2)(m_H^2+4m_W^2)C_0(m_Z^2,m_Z^2,s,m_W^2,m_W^2,m_W^2)\right]
  \\
  \Pi_1^{ZZ}(s)=&
  -\frac{\alpha_W}{16\pi m_W^2}\left[(4m_Z^2s+m_H^4+m_H^2m_Z^2-2m_Z^4)B_0(s,m_Z^2,m_Z^2)\right.\nonumber\\
    &\left.+(4m_H^2m_Z^2s+m_H^6-3m_H^2m_Z^4-6m_Z^6)C_0(m_Z^2,m_Z^2,s,m_Z^2,m_H^2,m_Z^2)\right]
  \\
  \Pi_1^{tt}(s)=&
  \frac{3\alpha_Wm_t^2}{8\pi m_W^2}\left[sB_0(s,m_t^2,m_t^2)+2m_t^2(s-2m_Z^2)C_0(m_Z^2,m_Z^2,s,m_t^2,m_t^2,m_t^2)\right]
  \\
  \Pi_1^{HH}(s)=&
  \frac{3\alpha_Wm_H^2}{16\pi m_W^2}\left[(m_H^2-m_Z^2)B_0(s,m_H^2,m_H^2)\right.\nonumber\\&
    \left.-(2m_Z^2s+m_H^4-2m_H^2m_Z^2-3m_Z^4)C_0(m_Z^2,m_Z^2,s,m_H^2,m_Z^2,m_H^2)\right] 
   \\
  \Pi_2^{WW}(s)=&
  \frac{\alpha_W}{16\pi m_W^2}\left[(m_H^4+4m_W^2m_H^2+12m_W^4)B_0(s,m_W^2,m_W^2)\right.\nonumber\\
    &\left.+4m_W^2(s-2m_Z^2)(m_H^2+4m_W^2)C_0(m_Z^2,m_Z^2,s,m_W^2,m_W^2,m_W^2)\right]
  \\
  \Pi_2^{ZZ}(s)=&
  \frac{\alpha_W}{32\pi m_W^2}\left[(3m_H^4+6m_H^2m_Z^2+8m_Z^4)B_0(s,m_Z^2,m_Z^2)\right.\nonumber\\
    &\left.+2(4m_H^2m_Z^2s+m_H^6-3m_H^2m_Z^4-6m_Z^6)C_0(m_Z^2,m_Z^2,s,m_Z^2,m_H^2,m_Z^2)\right]\\
  \Pi_2^{tt}(s)=&
  -\frac{3\alpha_Wm_t^4}{4\pi m_W^2}\left[2B_0(s,m_t^2,m_t^2)+(s-2m_Z^2)C_0(s,m_Z^2,m_Z^2,m_t^2,m_t^2,m_t^2)\right]\\
  \Pi_2^{HH}(s)=&
  \frac{3\alpha_Wm_H^2}{32\pi m_W^2}\left[(m_H^2+2m_Z^2)B_0(s,m_H^2,m_H^2)\right.\nonumber\\
    &\left.
    +2(2m_Z^2s+m_H^4-2m_H^2m_Z^2-3m_Z^4)C_0(m_Z^2,m_Z^2,s,m_H^2,m_Z^2,m_H^2)\right]
\end{flalign}
\end{widetext}
where $B_0$ and $C_0$ are the Passarino-Veltman functions \cite{'tHooft:1978xw}.
Note in the above equations, only the imaginary part of both sides will be used.

One can verify explicitly that the WI in Eq.~(\ref{eq:wi2}) is satisfied, \ie~with
the above definition, we should have
\begin{flalign}
  \Pi_{HH}^{(XX)}(s)=\Pi_{1}^{(XX)}(s)+\Pi_{2}^{(XX)}(s)
  \label{}
\end{flalign}
for $(WW)$, $(ZZ)$, $(tt)$, and $(HH)$, respectively.

\begin{figure}[ht]
  \begin{center}
    \psfig{figure=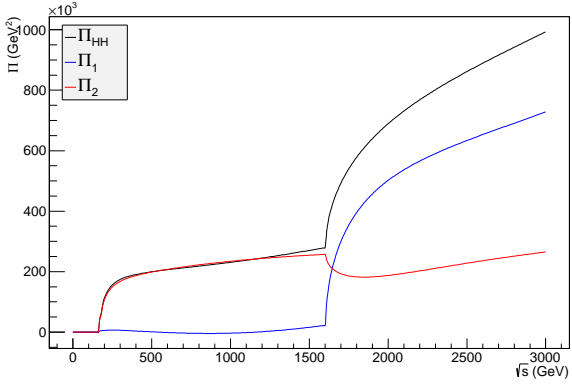,width=\columnwidth}
    \psfig{figure=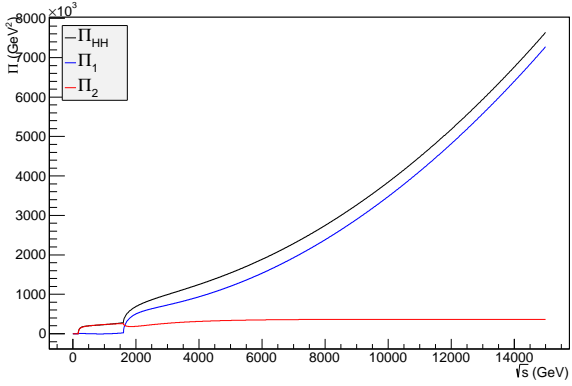,width=\columnwidth}
  \end{center}
  \caption{Comparison of $\Pi_{HH}$, $\Pi_1$ and $\Pi_2$.}
  \label{fig:sfos}
\end{figure}

Note that $\Pi_1$ corresponds to $HZZ$ and $\Pi_2$ corresponds to
$H\phi^0\phi^0$.  In \fig{fig:sfos}, we show a comparison of $\Pi_{HH}$,
$\Pi_1$ and $\Pi_2$.  At large energy, $\Pi_{HH}$ and $\Pi_1$ display the same
behaviour, while $\Pi_2$ (\ie~$H\phi^0\phi^0$) is negligible.  This justifies
the use of our operator ${\cal O}_\Pi$ in Eq.~(\ref{eq:op1}) at high energy,
as it generates the right $HVV$ vertex. At lower energy, however, $\Pi_2$
dominates and displays a ``bump'' above threshold. Because the operator given
in Eq.~(\ref{eq:op1}) does not give rise to the right $H\phi^0\phi^0$ vertex,
it is clear that had we chosen this operator in the EFT scheme, we would have
mistakenly considered this ``bump'' as part of $HVV$ vertex, resulting in an
unphysical excess at low energy.

To reproduce the right behaviour in both low and high energy regions we use the combined operator
\begin{equation}
  {\cal \bar{O}}_\Pi={\cal O}_{\Pi_1}+{\cal O}'_{\Pi_2}
  \label{eq:opgood}
\end{equation}
with $\Pi_1$ and $\Pi_2$ given by \eqn{eq:pi12}.

\begin{figure}[ht]
  \begin{center}
    \psfig{figure=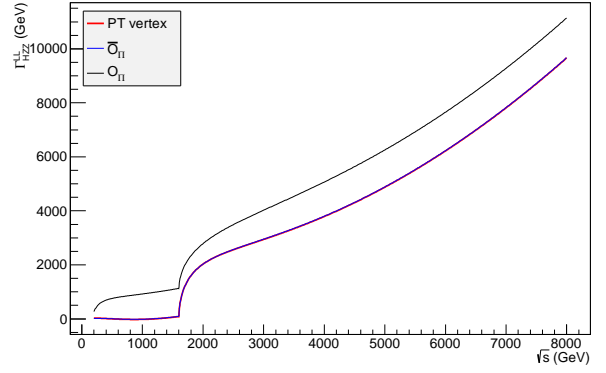,width=\columnwidth}
    \psfig{figure=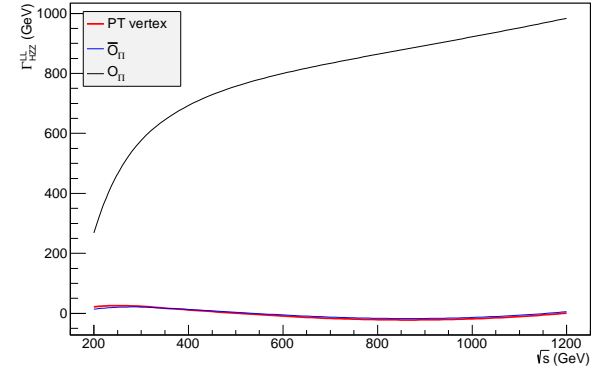,width=\columnwidth}
  \end{center}
  \caption{The longitudinal components of the $HZZ$ vertex.}
  \label{fig:hzzll}
\end{figure}

\begin{figure}[ht]
  \begin{center}
    \psfig{figure=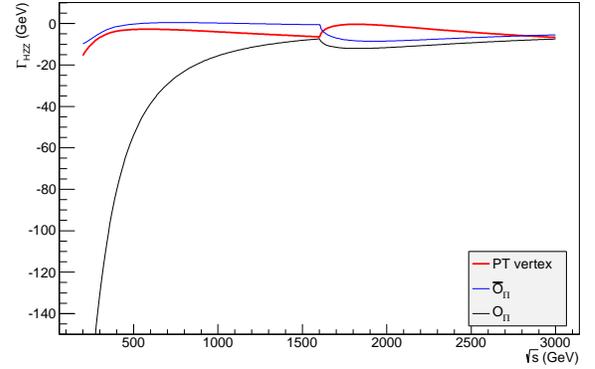,width=\columnwidth}
  \end{center}
  \caption{The transverse components of the $HZZ$ vertex.}
  \label{fig:hzztt}
\end{figure}

This operator gives a better description, as shown in
\fig{fig:hzzll} where we compare the longitudinal component of the $HZZ$
vertex derived from operator ${\cal O}_\Pi$ (\eqn{eq:op1}) and the improved
operator ${\cal \bar{O}}_\Pi$ (\eqn{eq:opgood}) with the actual calculation at
one loop with the PT. The improved operator describes the $HZZ$ vertex very
well.
Figure~\ref{fig:hzztt} compares the transverse component of the $HZZ$ vertex.
The improved operator comes closer to the actual transverse component.

When applying our scheme with the operator ${\cal \bar{O}}_\Pi$ in Eq.~(\ref{eq:opgood}), we prefer to work with the
complex-pole renormalization. To do this we need the following counterterms for
$\Pi_{HH}$:
\begin{equation}
  \Pi_{HH}^R(s)=\Pi_{HH}(s)-\delta s_H+(s-s_H)\delta Z
  \label{eq:piren}
\end{equation}
where $s_H$ is the pole. $\delta s_H$ and $\delta Z$ are given by
\begin{flalign}
  \delta s_H&=\Pi_{HH}(s_H)\\
  \delta Z&=-\Pi'_{HH}(s_H)
  \label{eq:deltas}
\end{flalign}
For the Feynman rules,
essentially we are going to replace all the $\Pi$'s in (\ref{eq:fr}) by $\Pi_1$,
and include a factor of $1+\delta Z$ for the SM $HVV$ vertex to account for the
wavefunction renormalization.

\section{Applications}

The treatment of the propagator of the Higgs  is of immediate  relevance for the LHC. 
As simple testing ground of our proposal and comparisons to the conventional methods, we consider three processes of particular phenomenological importance at the LHC for a scalar boson (which for brevity, we identify with an hypothetical heavy Higgs): vector boson scattering, $t\bar{t}$ production via vector boson fusion and Higgs production via gluon fusion. We have compared the effective approach described above in Eqs.~(\ref{eq:EFTmass}), (\ref{eq:EFTprop}) and (\ref{eq:fr}), with two other schemes: 
\begin{enumerate}
\item A naive inclusion of the self energy, \ie, using the following propagator
\begin{equation}\label{eq:naiveprop}
\frac{i}{\Delta_H(s)}=\frac{i}{s-s_H+\Pi_R(s)}\ ,
\end{equation}
without changing anything else. Here $s_H=\mu^2-i\mu\gamma$.

\item The CMS scheme, 
\begin{equation}\label{eq:CMSprop}
\frac{i}{\Delta_H(s)}=\frac{i}{s-s_H}\ .
\end{equation}
\end{enumerate}

A modified version of {\sc MadGraph}~\cite{Alwall:2011uj}, with the implementation of the effective Lagrangian approach and the naive propagator with the PT self energy is used to generate events. As
SM input parameters we take:
\begin{flalign}
&m_Z=91.188 {\rm GeV}\\
&G_F=1.16639\times 10^{-5} {\rm GeV}^{-2}\\
&\alpha^{-1}=132.507 \\
&m_t=173\,{\rm GeV}\,.
\end{flalign}
The pole mass is
\begin{equation}
\mu=800\,{\rm GeV}\,,
\end{equation}
and $\Pi_R(s)$ is the imaginary part of the PT self energy renormalized in the pole scheme. 
The factorization scale is set as the default dynamical scale of {\sc MadGraph} and the PDF set is CTEQ6l1~\cite{Pumplin:2002vw}.

\subsection{Vector Boson Scattering}

In $VV\ra VV$ scattering processes, the effective description allows one to
achieve a complete description of the Higgs line-shape at the resonance region
and at the same time it corrects the bad high-energy behavior originated from
the momenta dependent part of the self energy. As a bonus, we show  that
our definition avoids the need for including spurious $t$-channel widths which occur 
in the complex-mass scheme also affecting the high energy behavior of the scattering amplitudes. 

\begin{figure}[ht]
  \includegraphics[width=0.9\columnwidth,height=0.7\columnwidth]{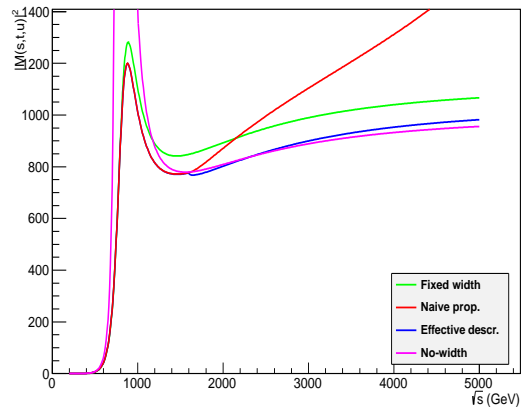}
\caption{$\sum_{hel} |M(ZZ\ra ZZ)|^2$ with scattering angle, $\theta=\pi/2$. The curves correspond to: fixed width scheme, \eqn{eq:CMSprop},
naive propagator, \eqn{eq:naiveprop}, the effective description and the no-width, in which the width is set to zero.}
\label{fig:amp_zz_cos0}
\end{figure}

\begin{figure}[ht]
\includegraphics[width=0.9\columnwidth,height=0.7\columnwidth]{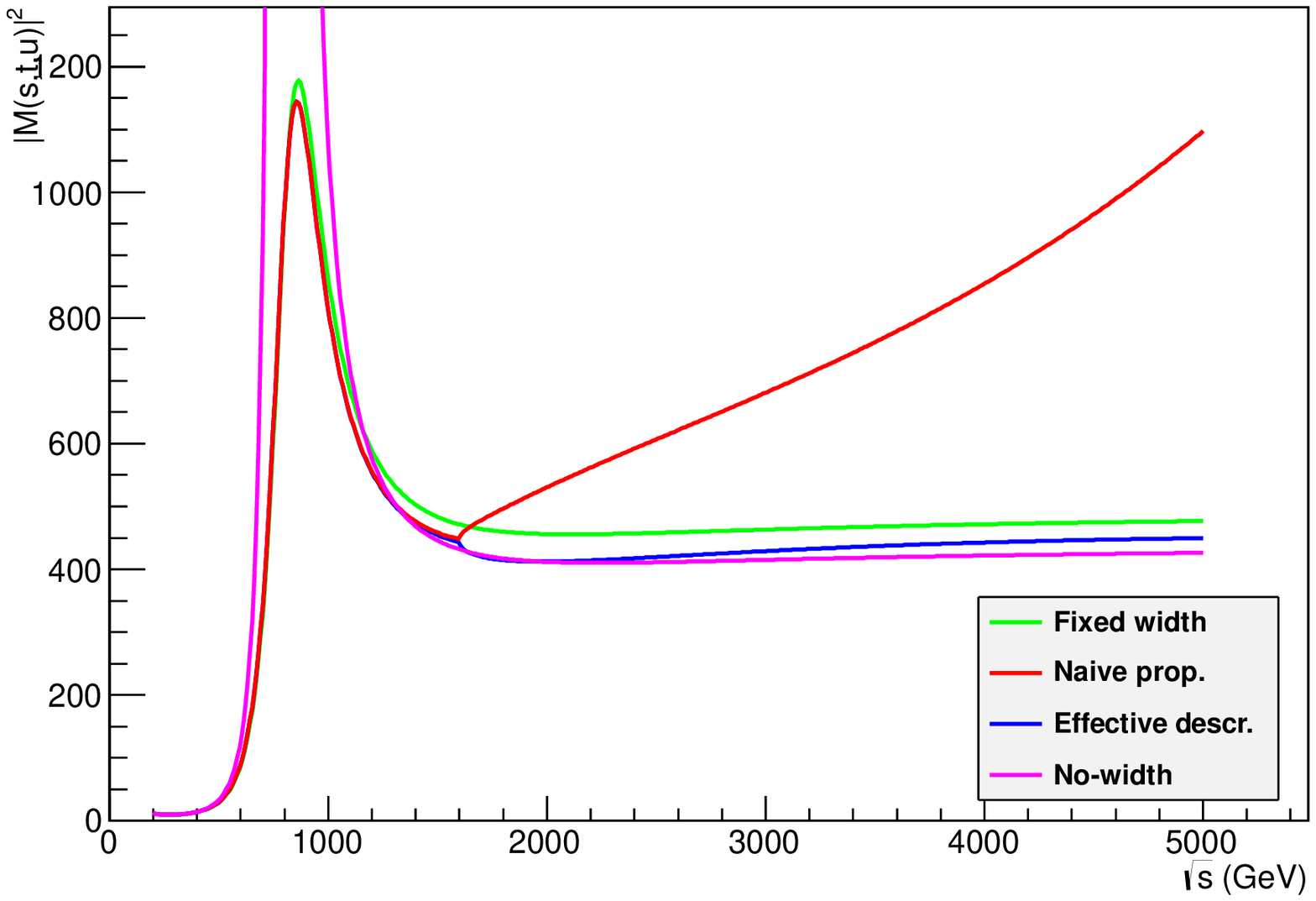}
\caption{$\sum_{hel} |M(W^+W^-\ra W^+W^-)|^2$ with scattering angle, $\theta=\pi/2$. The curves correspond to: fixed width scheme, \eqn{eq:CMSprop},
naive propagator, \eqn{eq:naiveprop}, the effective description and the no-width, in which the width is set to zero.}
\label{fig:amp_ww_cos0}
\end{figure}

In \fig{fig:amp_zz_cos0} we show the energy behaviour of the $ZZ\ra ZZ$ scattering amplitude summed over
helicities, $\sum_{hel}|M(s,t,u)|^2$, at scattering angle $\cos\theta=0$. The
fixed width scheme, \eqn{eq:CMSprop}, naive propagator, \eqn{eq:naiveprop}, the
effective description and a case in which the width is set to zero are
presented. The agreement between the effective scheme and
the naive propagator at the resonance region is pretty good. The difference with respect to the fixed
width scheme is evident. At high energy,  the naive propagator diverges, while the effective
description behaves correctly. Similar comments can be made about
$W^+W^-\ra W^+W^-$ amplitude, shown in \fig{fig:amp_ww_cos0}. 

\begin{figure}[ht]
\includegraphics[width=0.9\columnwidth,height=0.7\columnwidth]{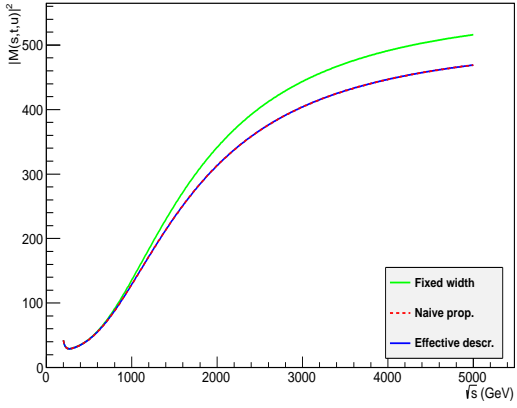}
\caption{$\sum_{hel} |M(W^+W^+\ra W^+W^+)|^2$ with scattering angle, $\theta=\pi/2$. The curves correspond to: fixed width scheme, \eqn{eq:CMSprop},
naive propagator, \eqn{eq:naiveprop} and the effective description. The no-width case is equivalent to the last two.}
\label{fig:amp_wpwp_cos0}
\end{figure}

The fact that in both $ZZ\ra ZZ$ and $W^+W^-\ra W^+W^-$ the fixed
width scheme differs from the effective approach at the high energy region
indicates that the spurious $t$-channel width gives a non-negligible
contribution. This fact can be verified by comparing the different schemes with
the no-width case. Moreover, in the case of  $W^\pm W^\pm\ra W^\pm W^\pm$, shown in \fig{fig:amp_wpwp_cos0}, the effective description and naive propagator are equivalent to the no-width case and the
excess observed in the amplitudes in the fixed width scheme comes from the
spurious width in the $t$ and $u$-channels.

At the LHC, the differences shown above may become important for a broad resonance. Despite the
fact that a light Higgs has been observed, there is still room for new heavy
and eventually broad resonances, e.g. in scaled-up QCD or in 2HDM's. The $VV$ scattering are
embedded in more complex processes of the form $qq\ra qqVV$, where the two final
state jets are emitted with high energy in the forward-backward region of the
detectors and the vector bosons decay into two fermions with high $p_T$ through
the central region. We study the processes $uc \ra uc ZZ$ and $us \ra dc
W^+W^-$ assuming the nominal energy of LHC, $E_{CM}=14\TeV$.

\begin{figure}[ht]
\includegraphics[width=0.9\columnwidth,height=0.7\columnwidth]{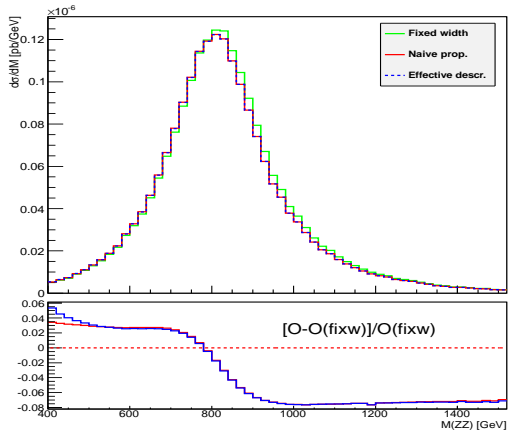}
\caption{Mass distribution of $ZZ$-system in the process $uc\ra ucZZ$ around the resonance peak. The cuts listed in the left column of \tbn{tab:vvcuts} have been applied.}
\label{fig:uc-uczz_mzz_genc}
\end{figure}

\begin{figure}[ht]
\includegraphics[width=0.9\columnwidth,height=0.7\columnwidth]{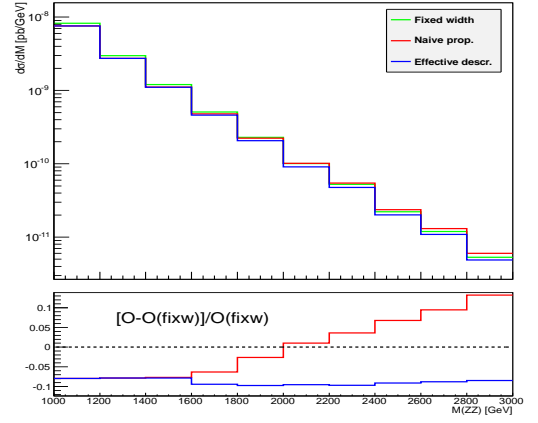}
\caption{Mass distribution of $ZZ$-system in the process $uc\ra ucZZ$ at the high energy region. All cuts listed in \tbn{tab:vvcuts} have been applied.}
\label{fig:uc-uczz_mzz_allp-m1000}
\end{figure}

\begin{figure}[ht]
\includegraphics[width=0.9\columnwidth,height=0.7\columnwidth]{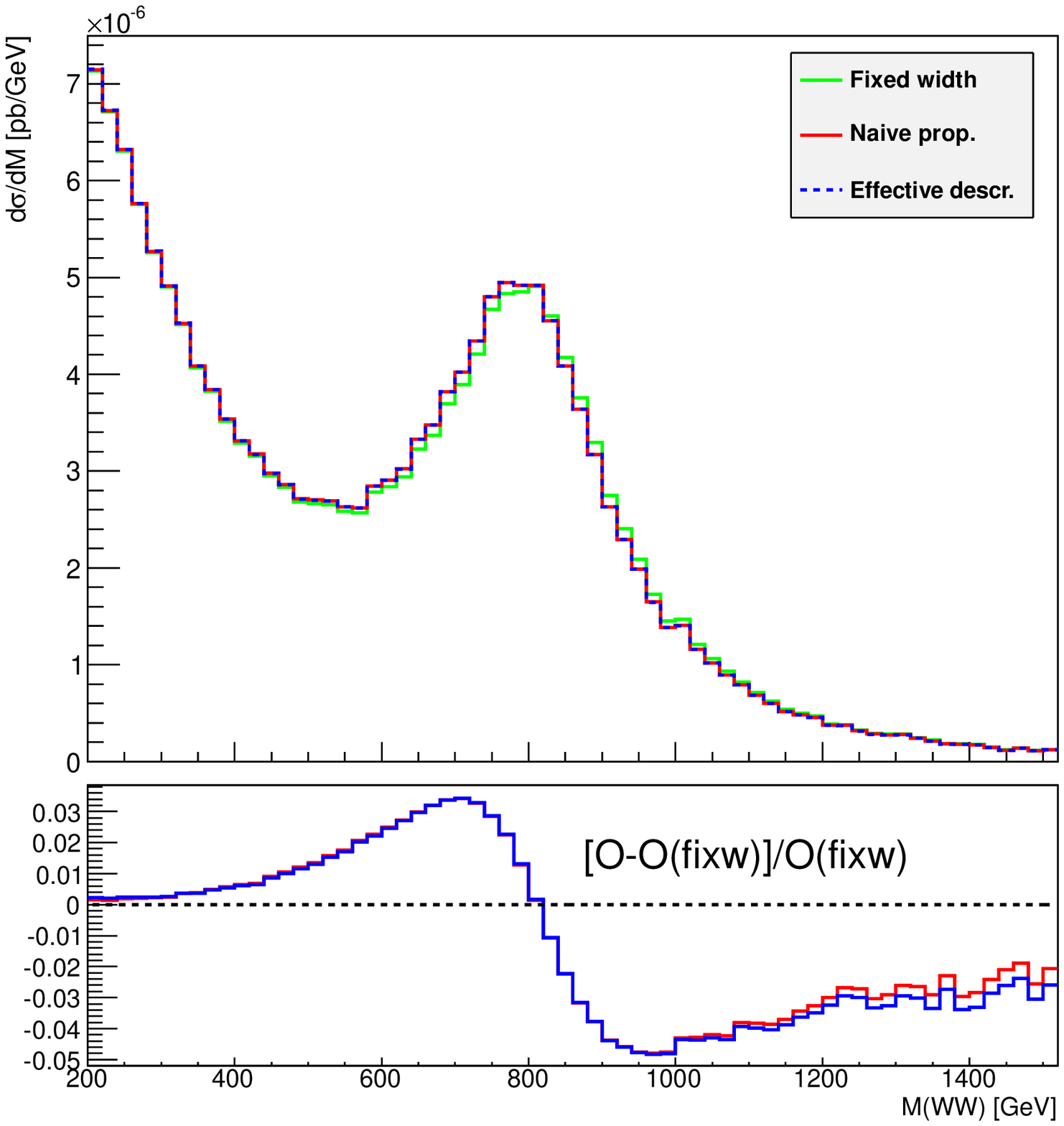}
\caption{Mass distribution of $WW$-system in the process $us\ra dcW^+W^-$ around the resonance peak. The cuts listed in the left column of \tbn{tab:vvcuts} have been applied.}
\label{fig:us-dcww_mww_genc}
\end{figure}

In \figsc{fig:uc-uczz_mzz_genc}{fig:uc-uczz_mzz_allp-m1000},  the
distribution of the invariant mass of the $ZZ$-system is shown. In
\fig{fig:uc-uczz_mzz_genc}, the resonant region is shown. A basic set of
selection cuts to enhance vector boson scattering contribution, listed in the
left column of \tbn{tab:vvcuts}, have been applied. The effective description
fits well with the running behavior of the Higgs propagator. In
\fig{fig:uc-uczz_mzz_allp-m1000}, the high-energy region is put in 
evidence. To better appreciate the differences between schemes at LHC energy, a
further set of cuts has been added (right column of \tbn{tab:vvcuts}). As
expected, the effective approach gives a well behaved distribution at such
energies contrary to the naive propagator and with a rate about 10\% lower than
the fixed width scheme. This difference amounts to the $t$-channel spurious
contribution present in the fixed width case. Similar conclusions can be drawn
from \figsc{fig:us-dcww_mww_genc}{fig:us-dcww_mww_allp-m1000}, where the
reconstructed $WW$-system invariant mass distribution for the $us\ra dcW^+W^-$
process is shown.  

\begin{figure}[ht]
\includegraphics[width=0.9\columnwidth,height=0.7\columnwidth]{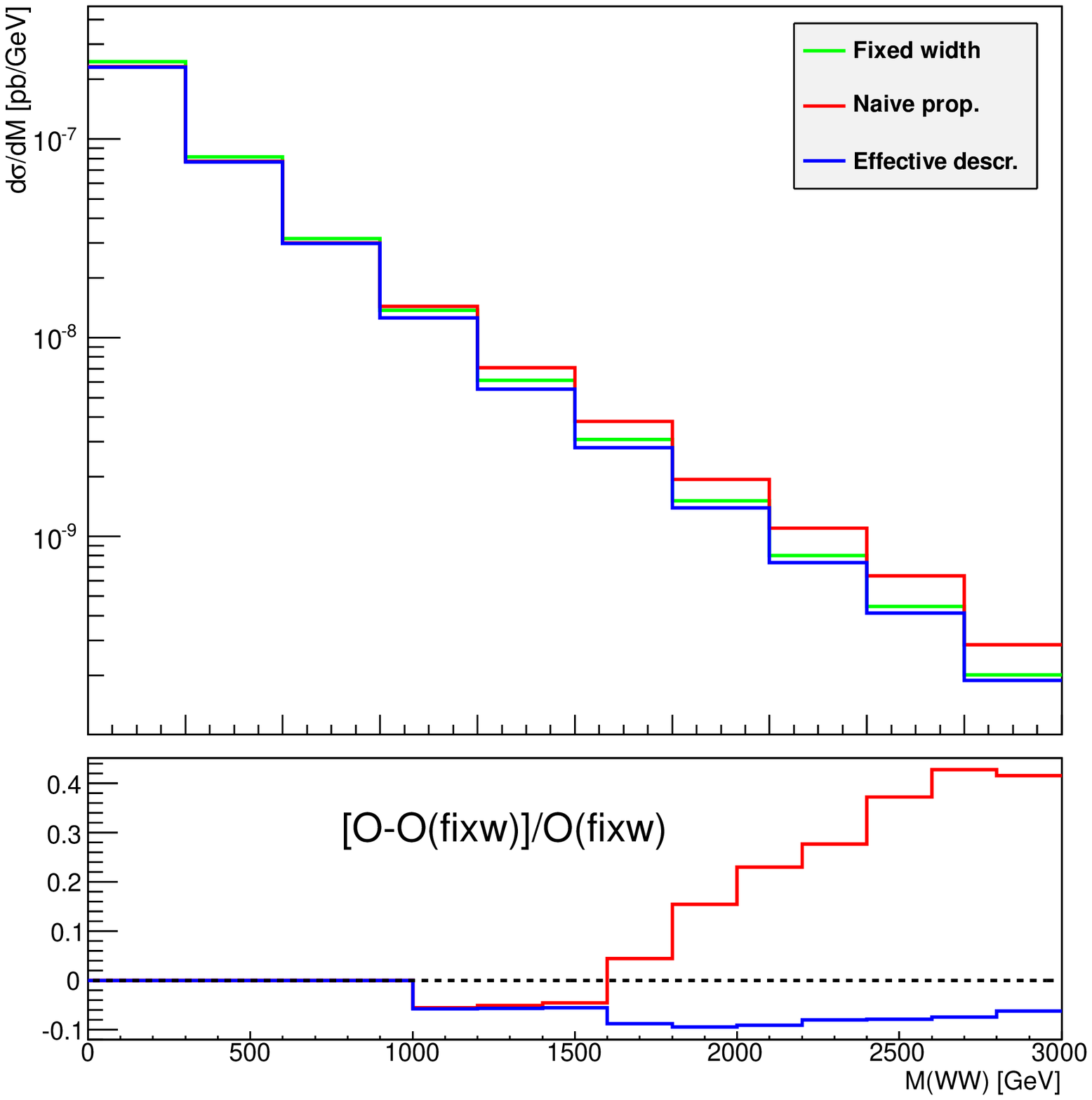}
\caption{Mass distribution of $WW$-system in the process $us\ra dcW^+W^-$ at the high energy region. All cuts listed in \tbn{tab:vvcuts} have been applied.}
\label{fig:us-dcww_mww_allp-m1000}
\end{figure} 

\begin{table}[h!]
\begin{tabular}{|c|c|}
\hline
Basic & Extra \\
\hline
$p_T(j)>10\,{\rm GeV}$ & $p_T(V)>400\,{\rm GeV}$\\
\hline
$2<\eta(j)<10$ 	 & $\Delta\eta(jj)>4.8$ \\
\hline
$\Delta R(j,j)>4$& $\eta(V)<2$ \\
\hline
$M(jj)>100\,{\rm GeV}$ & $M(jj)>1000\,{\rm GeV}$ \\
\hline
   		 & $\Delta \eta (V,j)>1$ \\
\hline
\end{tabular}
\caption{Cuts to enhance vector boson fusion. Basic ones on the left column and the extra ones on the right column.}
\label{tab:vvcuts}
\end{table}

\subsection{$W^+ W^+ \to H\ra t\bar{t}$ production}

\begin{figure}[t]
\includegraphics[width=0.9\columnwidth,height=0.7\columnwidth]{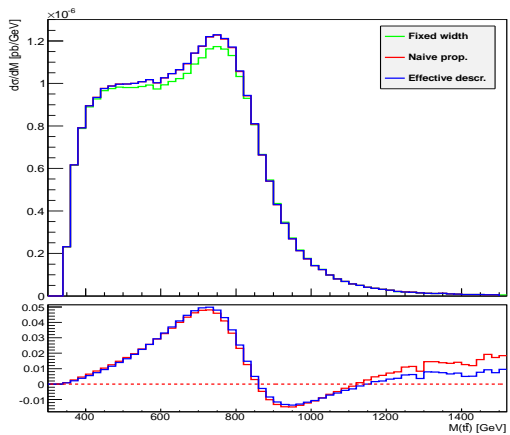}
\caption{Mass distribution of $t\bar{t}$-system in the process $us\ra dct\bar{t}$ around the resonance peak. The cuts listed in the left column of \tbn{tab:vvcuts} have been applied.}
\label{fig:mtt_genc}
\end{figure}

\begin{figure}[t]
\includegraphics[width=0.9\columnwidth,height=0.7\columnwidth]{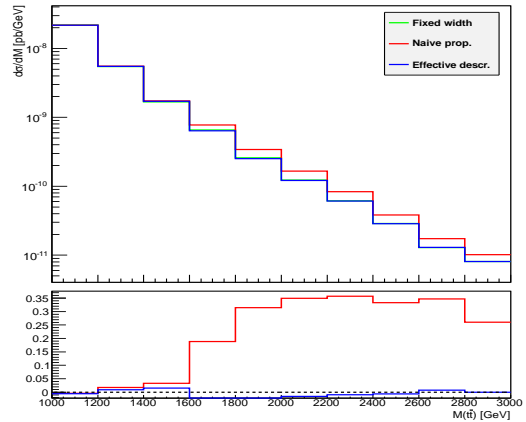}
\caption{Mass distribution of $t\bar{t}$-system in the process $us\ra dct\bar{t}$ at the high energy region. All cuts listed in \tbn{tab:vvcuts} have been applied.}
\label{fig:mtt_m1000_allp}
\end{figure}

In $t\bar{t}$ production, we can observe a similar behavior with respect to $ZZ\ra ZZ$ vector boson scattering. We have concentrated on the process $us\ra dct\bar{t}$, in which the Higgs in produced by $W^+W^-$ fusion and decayed to a pair of top quarks. The energy in the center of mass is set to $14$ TeV. In \fig{fig:mtt_genc}, the invariant mass distribution of $t\bar{t}$ at the resonant region is presented. The cuts shown in the left column of \tbn{tab:vvcuts} have been applied in order to enhance the vector boson fusion contribution. Here again, the effective description describes the functional form of the propagator, which can go up to 5\% of difference w.r.t the fixed width scheme. As seen in \fig{fig:mtt_m1000_allp}, in the high mass region, the effective description is dumped down by the effective $WWH$ vertex and does not grow with energy as is the case in which the naive propagator is adopted. The extra cuts shown in the right-hand column of \tbn{tab:vvcuts} have been added in order to highlight  the differences better.

\subsection{Gluon-gluon Fusion}

For the study of a heavy Higgs produced via gluon-gluon fusion and decayed to a $W$-boson pair, $gg\ra W^+W^-$, it is very important to consider the complete set of diagrams due to delicate gauge cancellations that control the high energy behavior. For this purpose we have relied on MCFM~\cite{MCFM} for evaluation of the matrix elements, taking into account all diagrams contributing at leading order (yet one loop) to the process $gg\rightarrow W^+W^-$, with $W$s decaying to leptons. 
Phase space integration and unweighted event generation have been carried out within the {\sc MadGraph} framework. The selection cuts shown in \tbn{tab:ggcuts} have been applied.

\begin{table}[h!]
\begin{tabular}{|c|}
\hline
 $p_T(\ell)>2\,{\rm GeV}$\\
\hline
 $\not{E}_T>2\,{\rm GeV}$ \\
\hline
 $\eta(\ell)<3$ \\
\hline
 $\Delta R(\ell\ell)>0.5$ \\
\hline
\end{tabular}
\caption{Cuts applied for $gg\rightarrow W^+W^-$ on the second one.}
\label{tab:ggcuts}
\end{table}

\begin{figure}[t]
\includegraphics[width=0.9\columnwidth,height=0.7\columnwidth]{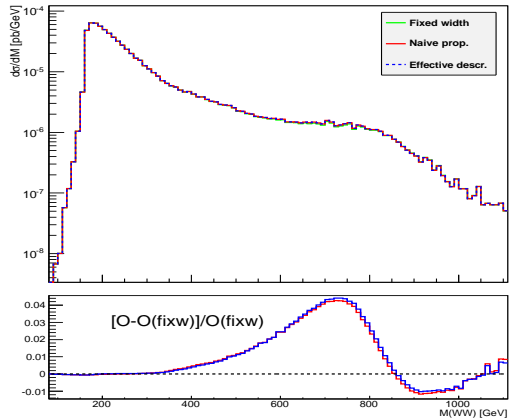}
\caption{Mass distribution of the reconstructed $WW$-system in the process $gg\ra W^+W^-\ra e^+\nu_e\mu^-\nu_\mu$, around the resonance peak. The cuts listed in \tbn{tab:ggcuts} have been applied.}
\label{fig:mww}
\end{figure}

\begin{figure}[t]
\includegraphics[width=0.9\columnwidth,height=0.7\columnwidth]{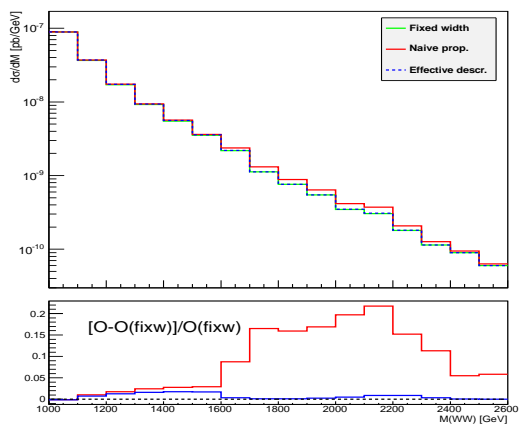}
\caption{Mass distribution of the reconstructed $WW$-system in the process $gg\ra W^+W^-$ at the high energy region. The cuts listed in \tbn{tab:ggcuts} have been applied.}
\label{fig:mww_m800}
\end{figure}

The mass distribution of $WW$-system in the three schemes considered if shown in \fig{fig:mww} and \fig{fig:mww_m800}. In \fig{fig:mww} we can see that the effective scheme and the naive propagator description present the same behavior around the resonance region, while the fixed width scheme shows a typical slighlity harder resonance. At high energy, the naive propagator diverges and the effective scheme and fixed width scheme are well behaved.

\section{Conclusions}
We have argued that it is possible to consistently and efficiently include running width effects for a 
heavy Higgs-like boson employing an EFT method.   We can summarize the main points of our approach as follows: 
\begin{itemize}
\item
Introducing a width for an unstable particle amounts to a rearrangement of the
perturbative expansion where the corrections to the two-point function are
resummed in the propagator.  The addition of the operator ${\cal \bar{O}}_\Pi$
defined in Eq.~(\ref{eq:opgood})  allows to effectively perform such resummation in
a gauge invariant and unitary way while keeping the full virtuality dependence
of the self-energy. We have shown that in the limit where such  dependence can
be neglected our scheme is equivalent to the CMS.
\item
At leading order, one has the freedom to choose the functional form of
$\Pi(s)$.  We propose to use the exact one-loop PT self-energy correction. The
rationale  is that such self-energies are gauge invariant and by exploiting the
WI's, we demand ${\cal \tilde{O}}_\Pi$ to mimic the most important one-loop
corrections as much as possible.  In practice, however, using any other form of
$\Pi(s)$ does not break either gauge invariance or unitarity. In particular,
one could avoid the need for a spurious non-zero width for $t$-channel
propagators.
\item 
EW higher-order corrections can be still performed in the CMS, without loss of accuracy or double counting issues.
In practice, one can include the running width effects via the EFT at the leading order and 
neglect the virtuality dependence at NLO, \ie, employ the usual CMS for the NLO term.
\end{itemize}
In conclusion, in this work we have considered the case of how to consistently define a running width in the case of a heavy SM Higgs. The same approach can be used, for example, in the context of a Two-Higgs-Doublet model and applied to the current searches for new scalar states at the LHC. Extension to gauge vectors and heavy fermion states, on the other hand, are not straightforward and 
need further investigation.

\section{Acknowledgements}
We would like to thank Giampiero Passarino for many discussions, patient
explanations, and comments on the manuscript.
We are grateful  to Vincent Mathieu for helpful discussion on the pinch technique.
The work of D.B.F has been supported by an FSR incoming post-doctoral fellowship
co-funded by the Marie Curie actions of the European Commission under contract
SPER/DST/266-1118906, by the BELSPO (by the IAP Program,
BELSPO VII/37, the IISN-FNRS convention 4.4511.10 ``Fundamental interactions").


\begin{thebibliography}{99}%

\bibitem{:2012an}
  G.~Aad {\it et al.}  [ATLAS Collaboration],
  arXiv:1207.0319 [hep-ex];
[CMS Collaboration],
CMS-PAS-HIG-12-020.

\bibitem{:2012gu} 
  S.~Chatrchyan {\it et al.}  [CMS Collaboration],
  Phys.\ Lett.\ B {\bf 716}, 30 (2012)
  [arXiv:1207.7235 [hep-ex]].

\bibitem{Kauer:2012hd}
  N.~Kauer and G.~Passarino,
  JHEP {\bf 1208} (2012) 116
  [arXiv:1206.4803 [hep-ph]].

\bibitem{Passarino:2012ri} 
  G.~Passarino,
  JHEP {\bf 1208}, 146 (2012)
  [arXiv:1206.3824 [hep-ph]].

\bibitem{Goria:2011wa} 
  S.~Goria, G.~Passarino and D.~Rosco,
  Nucl.\ Phys.\ B {\bf 864}, 530 (2012)
  [arXiv:1112.5517 [hep-ph]].

\bibitem{Argyres:1995ym} 
  E.~N.~Argyres, W.~Beenakker, G.~J.~van Oldenborgh, A.~Denner, S.~Dittmaier, J.~Hoogland, R.~Kleiss and C.~G.~Papadopoulos {\it et al.},
  Phys.\ Lett.\ B {\bf 358}, 339 (1995)
  [hep-ph/9507216].

\bibitem{Denner:1999gp} 
  A.~Denner, S.~Dittmaier, M.~Roth and D.~Wackeroth,
  Nucl.\ Phys.\ B {\bf 560}, 33 (1999)
  [hep-ph/9904472].

\bibitem{Denner:2005fg}
  A.~Denner, S.~Dittmaier, M.~Roth and L.~H.~Wieders,
  Nucl.\ Phys.\ B {\bf 724} (2005) 247
   [Erratum-ibid.\ B {\bf 854} (2012) 504]
  [hep-ph/0505042];
  A.~Denner and S.~Dittmaier,
  Nucl.\ Phys.\ Proc.\ Suppl.\  {\bf 160} (2006) 22
  [hep-ph/0605312].

\bibitem{Actis:2008uh} 
  S.~Actis, G.~Passarino, C.~Sturm and S.~Uccirati,
  Phys.\ Lett.\ B {\bf 669}, 62 (2008)
  [arXiv:0809.1302 [hep-ph]];
  S.~Actis, G.~Passarino, C.~Sturm and S.~Uccirati,
  Nucl.\ Phys.\ B {\bf 811}, 182 (2009)
  [arXiv:0809.3667 [hep-ph]];
  G.~Passarino, C.~Sturm and S.~Uccirati,
  Nucl.\ Phys.\ B {\bf 834}, 77 (2010)
  [arXiv:1001.3360 [hep-ph]].
   
\bibitem{Beenakker:1996kn}
  W.~Beenakker, G.~J.~van Oldenborgh, A.~Denner, S.~Dittmaier, J.~Hoogland, R.~Kleiss, C.~G.~Papadopoulos and G.~Passarino,
  Nucl.\ Phys.\ B {\bf 500} (1997) 255
  [hep-ph/9612260].

\bibitem{Abbott:1980hw}
  L.~F.~Abbott,
  Nucl.\ Phys.\ B {\bf 185} (1981) 189;
  L.~F.~Abbott, M.~T.~Grisaru and R.~K.~Schaefer,
  Nucl.\ Phys.\ B {\bf 229} (1983) 372;
  A.~Denner, G.~Weiglein and S.~Dittmaier,
  Nucl.\ Phys.\ B {\bf 440} (1995) 95
  [hep-ph/9410338];
  A.~Denner and S.~Dittmaier,
  Phys.\ Rev.\ D {\bf 54} (1996) 4499
  [hep-ph/9603341].


\bibitem{Stuart:1991xk}
  R.~G.~Stuart,
  Phys.\ Lett.\ B {\bf 262} (1991) 113;
  Phys.\ Rev.\ Lett.\  {\bf 70} (1993) 3193;
  A.~Aeppli, G.~J.~van Oldenborgh and D.~Wyler,
  Nucl.\ Phys.\ B {\bf 428} (1994) 126
  [hep-ph/9312212];
  H.~G.~J.~Veltman,
  Z.\ Phys.\ C {\bf 62} (1994) 35.

\bibitem{Seymour:1995qg}
  M.~H.~Seymour,
  Phys.\ Lett.\ B {\bf 354} (1995) 409
  [hep-ph/9505211].
  
\bibitem{Beenakker:1999hi}
  W.~Beenakker, F.~A.~Berends and A.~P.~Chapovsky,
  Nucl.\ Phys.\ B {\bf 573} (2000) 503
  [hep-ph/9909472].

\bibitem{Beenakker:2003va}
  W.~Beenakker, A.~P.~Chapovsky, AKanaki, C.~G.~Papadopoulos and R.~Pittau,
  Nucl.\ Phys.\ B {\bf 667} (2003) 359
  [hep-ph/0303105].

\bibitem{Beneke:2003xh}
  M.~Beneke, A.~P.~Chapovsky, A.~Signer and G.~Zanderighi,
  Phys.\ Rev.\ Lett.\  {\bf 93} (2004) 011602
  [hep-ph/0312331];
  M.~Beneke, A.~P.~Chapovsky, A.~Signer and G.~Zanderighi,
  Nucl.\ Phys.\ B {\bf 686} (2004) 205
  [hep-ph/0401002].

\bibitem{Valencia:1990jp}
  G.~Valencia and S.~Willenbrock,
  Phys.\ Rev.\ D {\bf 42} (1990) 853;
  G.~Valencia and S.~Willenbrock,
  Phys.\ Rev.\ D {\bf 46} (1992) 2247.


\bibitem{Sirlin:1991rt}
  A.~Sirlin,
  Phys.\ Lett.\ B {\bf 267} (1991) 240;
  S.~Willenbrock and G.~Valencia,
  Phys.\ Lett.\ B {\bf 259} (1991) 373;
  M.~Passera and A.~Sirlin,
  Phys.\ Rev.\ Lett.\  {\bf 77} (1996) 4146
  [hep-ph/9607253];
  B.~A.~Kniehl and A.~Sirlin,
  Phys.\ Lett.\ B {\bf 440} (1998) 136
  [hep-ph/9807545];
  Phys.\ Rev.\ Lett.\  {\bf 81} (1998) 1373
  [hep-ph/9805390];
  F.~Jegerlehner, M.~Y.~.Kalmykov and O.~Veretin,
  Nucl.\ Phys.\ B {\bf 641} (2002) 285
  [hep-ph/0105304];
  Nucl.\ Phys.\ B {\bf 658} (2003) 49
  [hep-ph/0212319].
    
\bibitem{Gambino:1999ai}
  P.~Gambino and P.~A.~Grassi,
  Phys.\ Rev.\ D {\bf 62} (2000) 076002
  [hep-ph/9907254];
  P.~A.~Grassi, B.~A.~Kniehl and A.~Sirlin,
  Phys.\ Rev.\ Lett.\  {\bf 86} (2001) 389
  [hep-th/0005149];
  P.~A.~Grassi, B.~A.~Kniehl and A.~Sirlin,
  Phys.\ Rev.\ D {\bf 65} (2002) 085001
  [hep-ph/0109228].
   
\bibitem{Kniehl:2002wn}
  B.~A.~Kniehl, C.~P.~Palisoc and A.~Sirlin,
  Phys.\ Rev.\ D {\bf 66} (2002) 057902
  [hep-ph/0205304].

\bibitem{Cornwall:1981zr}
  J.~M.~Cornwall,
  Phys.\ Rev.\ D {\bf 26} (1982) 1453.

\bibitem{Cornwall:1989gv}
  J.~M.~Cornwall and J.~Papavassiliou,
  Phys.\ Rev.\ D {\bf 40} (1989) 3474.

\bibitem{Papavassiliou:1989zd}
  J.~Papavassiliou,
  Phys.\ Rev.\ D {\bf 41} (1990) 3179.

\bibitem{Degrassi:1992ue}
  G.~Degrassi and A.~Sirlin,
  Phys.\ Rev.\ D {\bf 46} (1992) 3104.

\bibitem{Papavassiliou:1995fq}
  J.~Papavassiliou and A.~Pilaftsis,
  Phys.\ Rev.\ Lett.\  {\bf 75} (1995) 3060
  [hep-ph/9506417].

\bibitem{Papavassiliou:1996zn}
  J.~Papavassiliou and A.~Pilaftsis,
  Phys.\ Rev.\ D {\bf 54} (1996) 5315
  [hep-ph/9605385].
    
\bibitem{Papavassiliou:1995gs}
  J.~Papavassiliou and A.~Pilaftsis,
  Phys.\ Rev.\ D {\bf 53} (1996) 2128
  [hep-ph/9507246].
  
\bibitem{Papavassiliou:1997fn}
  J.~Papavassiliou and A.~Pilaftsis,
  Phys.\ Rev.\ Lett.\  {\bf 80} (1998) 2785
  [hep-ph/9710380].

\bibitem{Papavassiliou:1998pb}
  J.~Papavassiliou,
  hep-ph/9905328.
  
\bibitem{Papavassiliou:1997pb}
  J.~Papavassiliou and A.~Pilaftsis,
  Phys.\ Rev.\ D {\bf 58} (1998) 053002
  [hep-ph/9710426].

\bibitem{'tHooft:1978xw}
  G.~'t Hooft and M.~J.~G.~Veltman,
  Nucl.\ Phys.\ B {\bf 153} (1979) 365;
  G.~Passarino and M.~J.~G.~Veltman,
  Nucl.\ Phys.\ B {\bf 160} (1979) 151.

\bibitem{Denner:1994nn}
    A.~Denner, G.~Weiglein and S.~Dittmaier,
      Phys.\ Lett.\ B {\bf 333} (1994) 420
        [hep-ph/9406204].

\bibitem{Hashimoto:1994ct}
  S.~Hashimoto, J.~Kodaira, Y.~Yasui and K.~Sasaki,
      Phys.\ Rev.\ D {\bf 50} (1994) 7066
	[hep-ph/9406271].

\bibitem{Papavassiliou:1994yi}
  J.~Papavassiliou,
  Phys.\ Rev.\ D {\bf 51} (1995) 856
  [hep-ph/9410385].

\bibitem{Papavassiliou:1996fn}
  J.~Papavassiliou, E.~de Rafael and N.~J.~Watson,
  Nucl.\ Phys.\ B {\bf 503} (1997) 79
  [hep-ph/9612237].

\bibitem{Grinstein:2007mp} 
  B.~Grinstein, D.~O'Connell and M.~B.~Wise,
  Phys.\ Rev.\ D {\bf 77}, 025012 (2008)
  [arXiv:0704.1845 [hep-ph]].
  
\bibitem{Weinberg:1978kz}
  S.~Weinberg,
  Physica A {\bf 96} (1979) 327.

\bibitem{Alwall:2011uj}
  J.~Alwall, M.~Herquet, F.~Maltoni, O.~Mattelaer and T.~Stelzer,
  JHEP {\bf 1106} (2011) 128
  [arXiv:1106.0522 [hep-ph]].

\bibitem{MCFM}
J.~Campbell, R.~K.~Ellis, C.~Williams,
MCFM - Monte Carlo for FeMtobarn processes,
http://mcfm.fnal.gov/.

\bibitem{Pumplin:2002vw}
  J.~Pumplin, D.~R.~Stump, J.~Huston, H.~L.~Lai, P.~M.~Nadolsky and W.~K.~Tung,
  JHEP {\bf 0207} (2002) 012
  [hep-ph/0201195].

\end{thebibliography}
\end{document}